\documentclass[a4paper]{article}

\usepackage{INTERSPEECH2021}
\usepackage{booktabs}
\usepackage{multirow}
\DeclareCaptionLabelFormat{andtable}{#1~#2  \&  \tablename~\thetable}
\usepackage{enumitem}
\usepackage{caption}
\title{Comparing Supervised Models And Learned Speech Representations For Classifying Intelligibility Of Disordered Speech On Selected Phrases}
\name{Subhashini Venugopalan$^1$, Joel Shor$^1$, Manoj Plakal$^1$, Jimmy Tobin$^1$, Katrin Tomanek$^1$,\\ Jordan R. Green$^{2,3}$, Michael P. Brenner$^{1,3}$}
\address{
  $^1$Google Research, 
  $^2$MGH Institute of Health Professions USA, 
  $^3$Harvard University USA}
\email{\{vsubhashini,mbrenner\}@google.com}

\begin{document}

\maketitle
\begin{abstract}
  Automatic classification of disordered speech can provide an objective tool for identifying the presence and severity of a speech impairment. Classification approaches can also help identify hard-to-recognize speech samples to teach ASR systems about the variable manifestations of impaired speech. Here, we develop and compare different deep learning techniques to classify the intelligibility of disordered speech on selected phrases. We collected samples from a diverse set of 661 speakers with a variety of self-reported disorders speaking 29 words or phrases, which were rated by speech-language pathologists for their overall intelligibility using a five-point Likert scale. We then evaluated classifiers developed using 3 approaches: (1) a convolutional neural network (CNN) trained for the task, (2) classifiers trained on non-semantic speech representations from CNNs that used an unsupervised objective~\cite{TRILL}, and (3) classifiers trained on the acoustic (encoder) embeddings from an ASR system trained on typical speech~\cite{arun_rnnt}. We found that the ASR encoder’s embeddings considerably outperform the other two on detecting and classifying disordered speech. Further analysis shows that the ASR embeddings cluster speech by the spoken phrase, while the non-semantic embeddings cluster speech by speaker.
  Also, longer phrases are more indicative of intelligibility deficits than single words. 
\end{abstract}
\noindent\textbf{Index Terms}: atypical speech, classification, intelligibility, speech disorders

\section{Introduction}

Speech production requires the activation of dozens of muscles responsible for coordinating the respiratory, phonatory, resonatory, and articulatory speech subsystems. While speech disorders such as dysarthria are caused by neurological diseases such as amyotrophic lateral sclerosis (ALS) and Parkinson’s disease (PD), impaired speech can also manifest from a wide variety of conditions such as cleft lip and palate, hearing loss, or stuttering. Regardless of the underlying etiology, speech problems are currently assessed using subjective ratings scales and observations. More objective approaches are, however, needed to address pressing clinical needs such as improving early detection, diagnostic accuracy, and clinical monitoring~\cite{green2013bulbar,yunusova2016profiling}.
Speech analytics that leverage machine classification are particularly well suited to address these diagnostic challenges. Recent reports have demonstrated, for example, the added value of speech analytics for early detection of speech decline in ALS~\cite{allison2017diagnostic}, stratifying patients into fast or slow progressing groups~\cite{rong2020speech},  and documenting the response to drug interventions designed to slow the rate of disease progression \cite{green2018additional}. %

Automatic classification of impaired speech could also play a critical role in reducing the performance gap in automatic speech recognition (ASR) between typical and disordered speech~\cite{gupta2016pathological,benzeghiba2007automatic}.
Although ASR technologies could have a major impact on the quality of life in persons with severe speech impairments, these are the speakers for which ASR systems are most likely to fail~\cite{russis2019}. Although, some techniques can improve performance on these speakers~\cite{personalized_asr}, none so far have bridged the gap. Poor performance has largely been attributed to lack of sufficient diversity in the training data~\cite{moore2018whistle}. Hence, automatic detection of disordered speech can enable large scale collection of hard-to-recognize speech samples that can help teach ASR systems about the variable manifestations of severely impaired speech.

Long-Short-Term-Memory(LSTM)~\cite{lstm} based models have been extensively used in the context of speech recognition, while classification of non-semantic audio, disordered speech, and in particular dysarthric speech, is often based on convolutional neural network (CNN) models~\cite{liu2018gmm} or machine learning models trained on handcrafted acoustic features~\cite{hansen1998nonlinear,baghai2012automatic}. CNNs are often preferred  for general audio and dysarthria classification~\cite{janbakhshi2020automatic} since their training process is (i) more suited to overcome the issue of limited training data available for such tasks, and (ii) better capture acoustic features associated with the underlying disorder as opposed to focusing on long-range content. %
More specifically, when training CNN models on audio, signals are often split into short segments (in our case we use 960ms) that may or may not overlap, and the label for the entire sample is propagated to each of the segments, thus boosting training data. Further, such short segments often lack sufficient context to capture semantic content, but are better suited to capture acoustic features that may exhibit characteristics of the disorder. However, many of the short segments may contain silence or noise or not exhibit characteristics of the impairment and may still require longer context. With clever techniques to increase training samples, there have also been LSTM-based dysarthria classifiers~\cite{mayle2019diagnosing,kim2018dysarthric,millet2019learning} that have performed well.

In this work we compare aspects of these different and commonly used deep learning techniques and study them in the context of automatically detecting intelligibility deficits and classifying intelligibility of disordered speech. Specifically, we compare the performance of 3 types of models:
\begin{itemize}[leftmargin=*]
\itemsep0em 
    \item A CNN model similar to those used for audio classification~\cite{Hershey}, but directly trained to predict intelligibility ratings on our dataset of disordered speech samples.
    \item Classifiers trained on non-semantic speech representations~\cite{TRILL} learned by CNNs using an unsupervised objective. These have performed consistently well on several non-semantic speech tasks~\cite{TRILL}.
    \item Classifiers trained on representations extracted from the LSTM encoder of a production quality ASR model~\cite{arun_rnnt}. %
    In particular, we consider the embeddings extracted from the encoder portion alone since it captures the acoustic features.
\end{itemize}

We compare these models on a labeled dataset of 15,246 disordered speech samples from 661 participants speaking 29 selected phrases. We analyze performance on different classification subtasks and across speaker cohorts of different intelligibility classes. 
Further, we study representations from the different models to compare their ability to discriminate between semantic content and intelligibility. %
We also examined if specific phrases were more indicative of intelligibility deficits. %

\section{Automatic classification of intelligibility}
Early works on classification of disordered speech focused on handcrafted acoustic features~\cite{hansen1998nonlinear,baghai2012automatic}.
 CNNs~\cite{janbakhshi2020automatic} and LSTMs~\cite{millet2019learning} and other machine learning based models~\cite{kodrasi2020automatic} have also been used to classify speech disorders. However, previous works have 
been based on smaller sets of speakers or etiologies than that studied here. Such works have also focused largely on discriminating dysarthric speech from typical speech e.g., from speakers with Parkinson's or ALS~\cite{janbakhshi2020automatic,vasquez2017speaker,an2018automatic,vaiciukynas2017parkinson} or dysarthria from apraxia~\cite{kodrasi2020automatic}.

In this work we %
focus particularly on detecting and classifying {\sl intelligibility} of speech samples from speakers with a variety of self-reported speech disorders. Intelligibility measures how well speech is understood by a human listener \cite{stipancic2018minimally}. In our dataset, each individual speaker is scored for intelligibility on a five-point Likert scale by speech-language pathologists (SLPs). 
We consider 3 different approaches to training classifiers to discriminate intelligibility. 
Our first model uses a CNN  trained specifically to predict intelligibility. Our second approach also uses CNNs but pretrained with an unsupervised objective; we then build classifiers on embeddings extracted from the CNNs. Our third approach is LSTM based:  we use representations learned by the LSTM encoder of an ASR system and build classifiers on those representations.

\subsection{Convolutional neural network (CNN-ResNetish)}
Our first approach is a convolutional network trained specifically for the classification task. %
Our CNN model is based on the ResNetish variant of the standard ResNet-50 architecture used for large scale audio classification~\cite{Hershey}. The model is initialized with random weights. As is typical for CNNs for audio tasks, the model takes  the spectrogram of the audio waveform as input. The speech sample (one recording) is split into non-overlapping 960ms segments that are then decomposed with a short-time Fourier transform (STFT) applied to 25ms windows every 10ms. The resulting spectrogram is integrated into 64 mel-spaced frequency bins, and the magnitude of the bins are log-transformed to obtain 96x64 log-mel spectrogram windows of the input. We train a CNN model on these spectrograms with a logistic loss for each score class. During training, all segments of an input are given the same label as the underlying speech sample. The model outputs a probability distribution for each score class and for each segment. We use the ADAM optimizer with a learning rate of 3e-5, mini-batch size of 64 segments (selected randomly across training speech samples) and train for 100 epochs. We choose the best performing model on the validation set. Since the label distribution  across the score classes are different,we weight the loss for each segment inversely to the frequency of the class label at the segment level. During inference, we  aggregate the segment level scores by taking a mean across all spectrogram segments to get a probability distribution of the scores over the entire speech sample. The argmax of these scores determines the final class prediction.
    
\subsection{Non-semantic speech representations (TRILL)}
For our second approach, we use representations learned by a convolutional network that is trained with an unsupervised objective. Specifically, we consider the TRILL and TRILL-distilled models~\cite{TRILL}. The backbone architecture of TRILL is also based on the ResNetish~\cite{Hershey} CNN model, whereas the TRILL-distilled model is based on the more efficient MobileNet~\cite{mobilenet}. These models also take input spectrogram context-windows from 960ms segments of audio/speech. 

Unlike the fully-supervised training objective, the TRILL models are trained using an unsupervised triplet loss. They consider 3 segments, where 1 segment is the anchor, a second segment which comes from the same audio clip forms the positive example, and a third segment from a different audio clip constitutes the negative example. The model is trained with a hinge loss such that the distance ($L_2$) between the representations of the anchor and negative example is larger than the distance between the representations of the anchor and positive example by at least some margin. The TRILL models were trained on a subset of AudioSet~\cite{audioset} training set clips possessing the "speech" label. We consider the pre-ReLU output of the first 512-depth convolutional layer in ResNetish (referred to as TRILL layer19),  and the penultimate average pool layer in MobileNet (referred to as TRILL-distilled in ~\cite{TRILL}). TRILL layer19 yields a 12288-d embedding, and TRILL-distilled yields a 2048-d embedding. 

For each model, we extract embeddings for each 960ms segment of the speech sample, and consider the average-pooled (over time) embedding as the representation of the speech sample. We then train a  Logistic Regression model on the embeddings to predict intelligibility, using 
 multinomial loss (identical to \cite{TRILL}) to get a distribution over classes. Additionally, we also evaluate the performance of a balanced Random Forest with the embeddings as features, implemented by scikit-learn~\cite{scikit-learn}

\subsection{ASR system encoder representations (ASR-enc)}
Our third set of models are based on the representations from an LSTM encoder that models acoustic inputs in a production quality ASR system. The ASR system we consider is a recurrent neural network transducer (RNN-T)~\cite{graves2013speech} model. Specifically, the architecture is based on He et. al.~\cite{he2019streaming} and has been trained with modifications on more diverse acoustic data to improve long-form speech recognition~\cite{arun_rnnt}. 

The RNN-T model consists of an \textit{encoder} and a \textit{prediction} network that use LSTMs, and a \textit{joint network} that uses feed-forward layers. In this work, we only consider the \textit{encoder} which models the acoustic inputs. The \textit{encoder} consists of 8 unidirectional LSTM layers. The acoustic frontend is modeled using 128-dimensional log-mel filterbank energies computed on 32ms windows with a 10ms hop. Features from 4 contiguous segments are stacked, and then sub-sampled by a factor of 3. This is input to the encoder network LSTMs which have 2048 units each. The outputs are then projected down to 640 units after each layer. The encoder network also uses a time-reduction layer after the $2^{nd}$ LSTM layer, which stacks output features from 2 contiguous timesteps and subsamples them by a factor of 2; the final encoder outputs are 640-d and at a 60ms frame rate.

Similar to our treatment of the TRILL models, we consider the average-pooled (over time) embeddings as the final representation of the speech sample, and train logistic regression and random forest models on the embedding features to predict class scores.

\section{Experiments}

We now describe the dataset, the classification tasks and objectives, and the evaluation metrics used in our experiments.

\subsection{Dataset of speakers and selected phrases}
\begin{figure}[!thb]
\vspace{-0.4cm}
  \centering
  \includegraphics[width=0.9\linewidth]{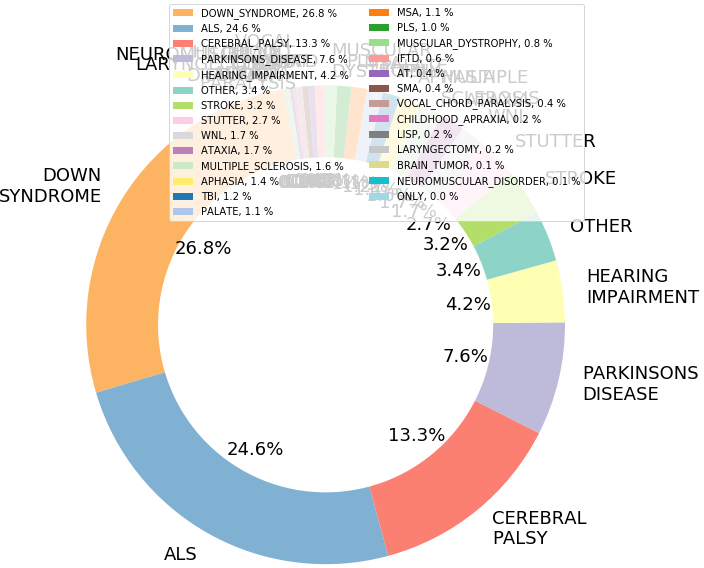}
  \caption{Distribution of etiologies in the dataset.}
  \label{fig:etiology_dist}
  \vspace{-0.8cm}
\end{figure}

\begin{figure}[!thb]
\vspace{-0.5cm}
  \centering
  \includegraphics[width=\linewidth]{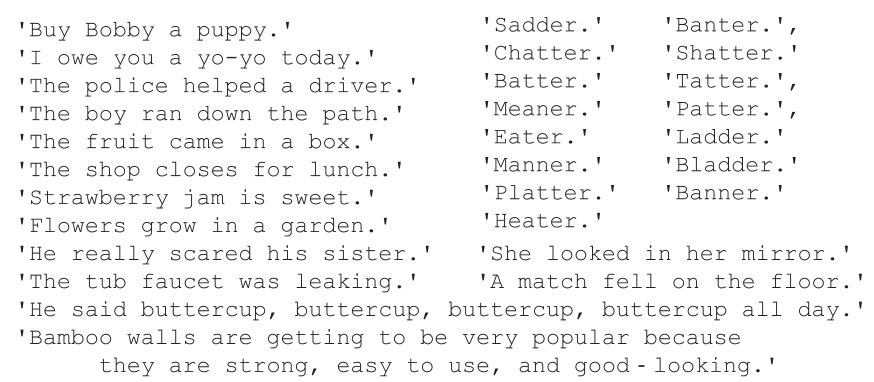}
  \caption{List of 29 words and phrases recorded by speakers.}
  \label{fig:phrase_list}
\vspace{-0.1cm}
\end{figure}

\begin{figure}[!thb]
\vspace{-0.2cm}
  \centering
  \includegraphics[width=\linewidth]{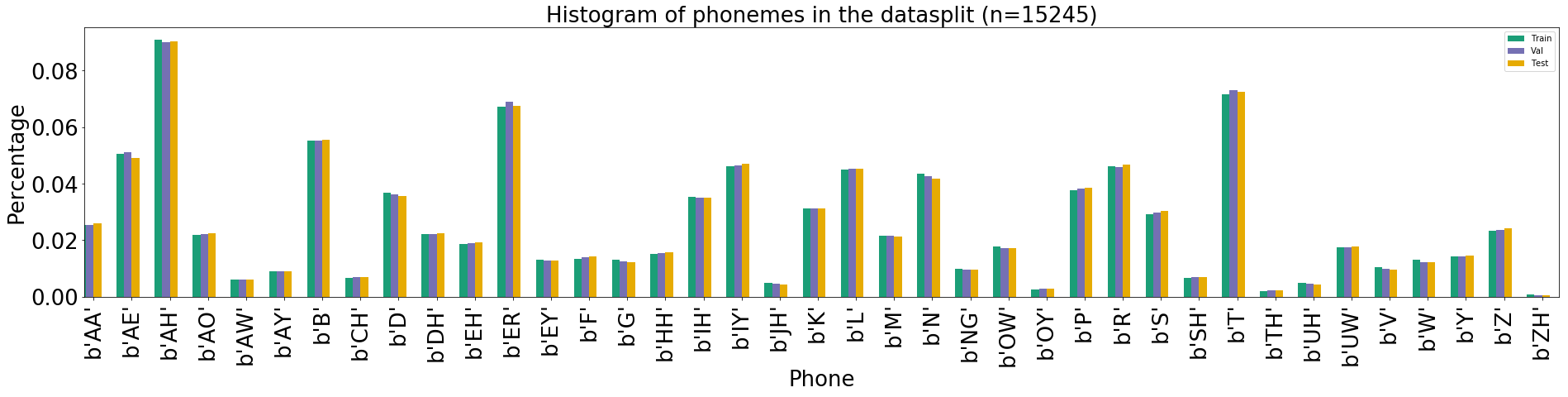}
  \caption{Distribution of phonemes in the datasplits. $x$-axis represents phonemes and $y$-axis the percentage. (Best viewed in high resolution)}
  \label{fig:phoneme_dist}
\vspace{-0.5cm}
\end{figure}

Our dataset is a human-rated small subset of \cite{euphonia1Mdata}. We use data from 661 speakers with a diverse set of self-reported speech disorders depicted in~Fig.~\ref{fig:etiology_dist}, including Down syndrome (26.8\%), ALS (24.6\%), Cerebral palsy (13.3\%), Parkinson's disease (7.6\%), hearing impairments (4.2\%) and others. The speakers each record 29 distinct words/phrases (Fig.~\ref{fig:phrase_list}) that have a mixed distribution of phonemes (Fig.~\ref{fig:phoneme_dist}). Speech Language Pathologists (SLPs) listened to the recordings for each speaker and assessed the overall intelligibility of the speaker on a five-point Likert scale. The scale was mapped to 5 classes - \textit{typical, mild, moderate, severe,} and \textit{profound}. %
We then split the speakers randomly into training, validation (val.) and test sets, with a distribution of 70:15:15. All our models were trained on the same splits. The number of speakers and utterances in each split for each label, along with the overall count is shown in Tab.~\ref{tab:split-stats}.
\begin{table}[!htb]
\setlength{\tabcolsep}{2pt}
\begin{center}
\begin{tabular}{l|ccc|ccc}
\toprule
\multirow{2}{*}{{Intelligibility}} & \multicolumn{3}{c|}{{\# speakers}} & \multicolumn{3}{c}{{\# utterances}} \\
 & \multicolumn{1}{|c}{{Train}} & \multicolumn{1}{c}{{Val.}} & \multicolumn{1}{c}{{Test}} &
\multicolumn{1}{|c}{{Train}} & \multicolumn{1}{c}{{Val.}} & \multicolumn{1}{c}{{Test}} \\
\midrule
TYPICAL &  160 &   30 & 23 & 3,875 & 734 & 544 \\
MILD    &  153 &   35 & 36 & 3,343 & 817 & 788 \\
MODERATE&   87 &   25 & 18 & 1,969 & 567 & 471 \\
SEVERE  &   54 &   12 & 14 & 1,113 & 316 & 388 \\
PROFOUND&   10 &    1 &  3 &   224 &   9 &  87 \\
\midrule
OVERALL &  464 &  103 & 94 & 10,524 & 2,443 & 2,278 \\
\bottomrule
\end{tabular}
\caption[splits]{Count of speakers and utterances in the data splits.}\label{tab:split-stats}
\end{center}
\vspace{-1cm}
\end{table}

\subsection{Classification objectives}
We train all our models on 4 different classification tasks based on the intelligibility ratings. Since the ratings are for each speaker, all utterances from a speaker have the same rating.

\textbf{5-class} This task is the 5-way classification task of labeling each utterance with one of the 5 ratings of \textit{0-typical, 1-mild, 2-moderate, 3-severe,} or \textit{4-profound}.

\textbf{3-class}  This task is a 3-way classification task where \textit{typical} and \textit{mild} are each a separate class of their own, but ratings of \textit{moderate, severe} and \textit{profound} are grouped into a single class. Hence, the label for each utterance corresponds to one of 3 ratings \textit{ 0-typical, 1-mild, 2-(moderate, severe,} or \textit{profound)}.


\textbf{2-class MODR$+$}  This task is a 2-way classification task where \textit{typical} and \textit{mild} are grouped into a single class and the ratings of \textit{moderate, severe} and \textit{profound} are grouped into another class. So each utterance maps to one of 2 labels for ratings \textit{ 0-(typical, mild)}, and \textit{1-(moderate, severe,} or \textit{profound)}.

\textbf{2-class MILD$+$} This essentially boils down to a 2-way classification task where we simply ask if the speech sample is \textit{typical} or not (i.e., disordered). So \textit{typical} is a class of its own and \textit{mild, moderate, severe} and \textit{profound} are grouped into the other class. So the label for each utterance maps to one of 2 ratings \textit{ 0-(typical)}, and \textit{1-(mild, moderate, severe,} or \textit{profound)}.

\vspace{-0.25cm}
\subsubsection{Evaluation metrics.}
\vspace{-0.2cm}
We report performance of our models on 3 evaluation metrics.

\noindent\textbf{1-vs-rest AUC} (AUC) This computes the Area Under the Receiver Operating Characteristic Curve for each class against the rest. This treats the classification task as a multi-label problem. We report the mean AUC of the different classes.

\noindent\textbf{F1 score} (F1) This is the balanced F-score which can be interpreted as the weighted average of the precision and recall of the model.

\noindent\textbf{Accuracy} (Acc.) Accuracy simply measures the number of correctly predicted samples over the total number of samples.

\vspace{-0.25cm}
\subsubsection{Pre-processing notes.}
\vspace{-0.2cm}
For all models we re-sample recordings to 16 kHz mono audio.

\vspace{-0.2cm}
\section{Results}
\begin{table*}[!tbh]
\vspace{-0.5cm}
\begin{center}
\begin{tabular}{l|ccc|ccc|ccc|ccc}
\toprule
\multicolumn{1}{c|}{{}} & \multicolumn{3}{c|}{{\textbf{2-class MILD$+$}}} & \multicolumn{3}{c|}{{\textbf{2-class MODR$+$}}} & \multicolumn{3}{c|}{{\textbf{3-class}}} &
\multicolumn{3}{c}{{\textbf{5-class}}} \\
\multicolumn{1}{c|}{{Models}} & 
AUC & F1 & Acc.  & AUC & F1 & Acc. & AUC & F1 & Acc. & AUC & F1 & Acc.\\
\midrule
CNN-ResNetish &  %
0.761 & 0.761  & 0.757 &
0.706 & 0.671  & 0.673 &
0.679 & 0.499  & 0.498 &
0.677 & 0.341  & 0.405 \\
TRILL (layer 19) &
0.723 & 0.708 & 0.709 &
0.631 & 0.643 & 0.650 &
0.638 & 0.445 & 0.448 &
0.603 & 0.373 & 0.387 \\
TRILL-distilled &
0.717 & 0.729 & 0.732 &
0.654 & 0.645 & 0.654 &
0.627 & 0.443 & 0.444 &
0.582 & 0.367 & 0.381 \\
ASR-enc &
\textbf{0.820} & \textbf{0.776} & \textbf{0.776} &
\textbf{0.812} & \textbf{0.754} & \textbf{0.763} &
\textbf{0.749} & \textbf{0.544} & \textbf{0.544} &
\textbf{0.771} & \textbf{0.448} & \textbf{0.459} \\
\bottomrule
\end{tabular}
\caption[qnt-eval]{
We report the mean 1-vs-rest AUC values, F1 score, and accuracy (Acc.) for different models on different classification objectives. Higher is better. \textbf{bold} indicates highest value.
}\label{tab:qnt-eval}
\end{center}
\vspace{-0.9cm}
\end{table*}

\begin{figure*}[!thb]
  \centering
  \includegraphics[width=\linewidth]{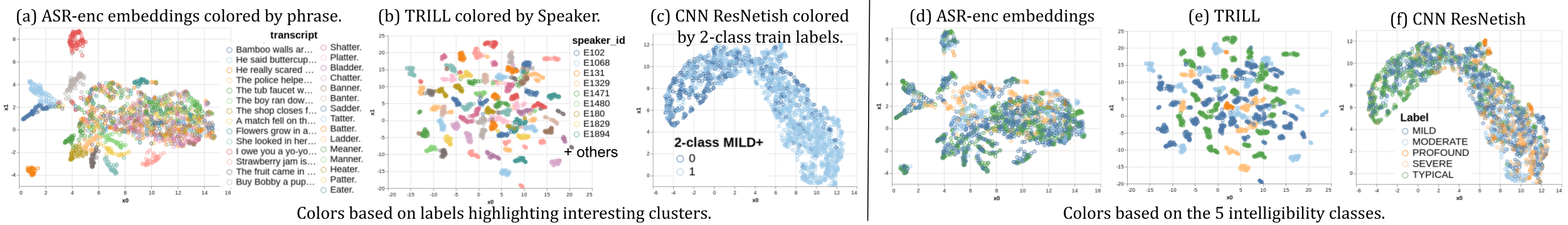}
  \caption{UMAP clusters of model representations. Samples are colored based on specific labels to highlight interesting clusters: (a) ASR embeddings cluster by phrase, (b) TRILL clusters by speakers, and (c) CNN-ResNetish clusters on training class labels. (d), (e) and (f) depict coloring based on the 5 intelligibility class labels for the 3 representations. (Best viewed in color and high-res.)}
  \label{fig:all_embs_viz}
  \vspace{-0.45cm}
\end{figure*}
The performance of all our models across the different classification tasks is reported in Tab.~\ref{tab:qnt-eval}. For the TRILL and ASR-enc-embeddings based models, we only show performance of the logistic regression model, which was consistently better than random forest for all representations and all classification tasks.

\textbf{ASR-enc representations perform best.} Surprisingly, the simpler classification models trained on the representations from the ASR \textit{encoder} achieve the best performance across all tasks. While we might expect the supervised CNN (CNN-ResNetish) model trained explicitly for the task to perform well, its performance is a distant second, particularly in terms of mean AUC. The models based on ASR representations also  maintain a high mean AUC on the 3-class and 5-class tasks.  However, in all tasks and for all models, there appears to be much room for improvement, since the best scores are  around 0.8 for the 2-way classification tasks.

\textbf{2-way tasks are simpler for all models.}
All models perform substantially better on the 2-class MILD$+$ and 2-class MODR$+$ classification tasks than the more fine-grained 3-class and 5-class classification tasks. Also, between the MILD$+$ and MODR$+$ tasks, models consistently perform better on the MILD$+$ task. One might have expected that the task of discriminating \textit{moderate} or above rating (MODR$+$) is easier than the (MILD$+$) task, since distinguishing mildly impaired speech from typical speech is more nuanced than identifying speech that is considerably impaired. The result indicates that the models can detect mild disordered speech, but this might  be an indicator that the range of mild intelligibility scored by the speech-language pathologists is quite large.

\vspace{-2pt}
\section{Analysis}

\subsection{Performance by intelligibility group}
\vspace{-1pt}
Table~\ref{tab:grp-perf} compares the performance of the 2-class MILD$+$ classifiers based on ASR encoder embeddings, TRILL (layer 19) and the CNN-ResNetish model grouped by intelligibility class labels. Utterances from speakers labeled moderate, profound or severe are easier to discriminate, and those labeled typical and mild are the hardest to discriminate.
\begin{table}[!htb]
\vspace{-0.25cm}
\setlength{\tabcolsep}{2pt}
\begin{center}
\begin{tabular}{l|cc|cc|cc}
\toprule
\multirow{2}{*}{{2-class MILD$+$}} & \multicolumn{2}{c|}{{ASR-enc-embs.}} & \multicolumn{2}{c|}{{TRILL (l.19)}} &
\multicolumn{2}{c}{{CNN-ResNetish}}\\
 & \multicolumn{1}{c}{{F1}} & \multicolumn{1}{c|}{{Acc.}} &
   \multicolumn{1}{c}{{F1}} & \multicolumn{1}{c|}{{Acc.}} &
   \multicolumn{1}{c}{{F1}} & \multicolumn{1}{c}{{Acc.}} \\
\midrule
TYPICAL &  0.766 &   0.621 & 0.659 & 0.491 & 0.706 & 0.546 \\
MILD    &  0.860 &   0.755 & 0.880 & 0.786 & 0.882 & 0.788 \\
MODERATE&  0.981 &   0.963 & 0.944 & 0.894 & 0.899 & 0.817 \\
SEVERE  &  0.959 &   0.921 & 0.904 & 0.825 & 0.934 & 0.876 \\
PROFOUND&  1.000 &   1.000 & 0.977 & 0.954 & 0.976 & 0.954 \\
\bottomrule
\end{tabular}
\caption[perf-sev]{Performance of 2-class MILD$+$ models grouped by intelligibility class. Scores increase with severity. Typical and mild are hardest to discriminate.}\label{tab:grp-perf}
\end{center}
\vspace{-1.2cm}
\end{table}

\vspace{-5pt}
\subsection{Longer phrases are more indicative of intelligibility}
\vspace{-1pt}
Next, we examine the performance at the phrase level for the best model i.e., ASR-enc-embeddings logistic regression model on the 2-class MILD$+$ task. The top two phrases with highest AUC for the model are: (1) ``He said buttercup, buttercup, buttercup, buttercup all day.''	 This had (AUC: 0.901,	F1: 0.849, accuracy=0.851), and (2) ``Bamboo walls are getting to be very popular because they are strong, easy to use, and good‐looking.'' (AUC:0.88, F1: 0.793, accuracy: 0.8). In contrast, the phrases with the lowest scores consist of single words, with the lowest scoring words  (1) ``Tatter.'' (AUC: 0.624, F1:	0.661, accuracy:	0.671), and (2) ``Chatter.'' (AUC: 0.764,	F1: 0.744, accuracy: 0.746). Irrespective of the actual content longer phrases had better performance than single words, indicating that more signal/context could help.

\vspace{-5pt}
\subsection{Comparing representations on speaker, intelligibility and content}
\vspace{-4pt}
Finally, Figure ~\ref{fig:all_embs_viz} shows clusters using UMAP~\cite{mcinnes2018umap} on representations from TRILL (layer 19), ASR-enc embeddings, and CNN-ResNetish (from the 2048-d fully connected layer of the 2-class MILD$+$ model).
In the ASR-enc based embeddings (Fig.~\ref{fig:all_embs_viz}a), longer sentences have clearer clusters, whereas single words are more spread out. In Fig.~\ref{fig:all_embs_viz}d, we can observe that utterances by speakers labeled severe and profound (orange, yellow) tend to cluster closer towards center, likely because of their poor intelligibility. In contrast, the TRILL  embeddings (Fig.~\ref{fig:all_embs_viz}b) cluster based on the identity of the speaker, and not so much by content (not depicted) or intelligibility class (Fig.~\ref{fig:all_embs_viz}e). The CNN-Resnetish model's embedding (Fig.~\ref{fig:all_embs_viz}c) appears to cluster more based on the training labels (here on the 2-class MILD$+$ labels) and doesn't seem to have any observable pattern on the full scale of 5 classes (Fig.~\ref{fig:all_embs_viz}f).

\vspace{-2pt}
\section{Conclusion}
\vspace{-1pt}
To conclude, we have demonstrated that deep learning models can classify speech samples from impaired speakers, with classifiers based on embeddings from state of the art ASR models outperforming task specific CNNs.  The models are able to classify the degree of intelligibility in speakers from a wide range of etiologies. Future directions include comparing the performance of these models to SLPs, as well as assess whether such classification can be carried out on more general speech. On the modeling side, we wish to consider alternative learned frontends (e.g., LEAF~\cite{zeghidour2021leaf}) and representations from  wav2vec~\cite{schneider2019wav2vec}, as well as finetuning the speech representations from TRILL, or the ASR systems on atypical speech training data.

\vspace{4pt}
\noindent \textbf{Acknowledgements} 
This study would not have been possible without the contributions and efforts of the hundreds of speakers who consented and provided their speech samples through g.co/euphonia. We thank Katie Seaver for assessing and labeling a portion of the speech samples, Aren Jansen for advice on the CNN-ResNetish model, Shanqing Cai and Dick Lyon for reviews on a draft of this work, and members of Team Euphonia for their data collection effort and feedback.

\newpage
\bibliographystyle{IEEEtran}

\bibliography{mybib}

\end{document}